\documentstyle[prb,aps,multicol,graphicx]{revtex}
	
\newcommand{\crossprod}{\times}

\begin{document}

\draft

\title{
Butterfly-like spectra and collective modes
of antidot superlattices in magnetic fields
}

\author{Egidijus Anisimovas\cite{email1} and Peter Johansson\cite{email2}}
\address{Department of Theoretical Physics, University of Lund, 
S{\"o}lvegatan 14 A, S-223 62 Lund, Sweden}

\date{\today}

\maketitle

\begin{abstract}
We calculate the energy band structure for  electrons in
an external periodic potential combined with a perpendicular magnetic field.
Electron-electron interactions are included within a Hartree approximation.
The calculated energy spectra display a considerable degree of 
self-similarity, just as the ``Hofstadter butterfly.''
However, screening affects the butterfly, most importantly
the bandwidths oscillate with magnetic field in a characteristic way.
We also investigate the dynamic response of the electron system
in the far-infrared (FIR) regime.
Some of the peaks in the FIR absorption spectra can be interpreted 
mainly in semiclassical terms, 
while others originate from inter(sub)band transitions.

\end{abstract}



\begin{multicols}{2}

Recent years have witnessed a considerable amount of research 
effort directed towards understanding of the physics of two-dimensional
electron systems (2DES) whose dimensionality is further restricted by
man-made periodic potentials and perpendicular magnetic fields. 
These include quantum dot arrays and antidot superlattices. 
Concentrating on the latter ones, one distinguishes two principal 
directions of experimental work: transport studies and far infrared 
(FIR) spectroscopy.
Some of the transport measurements\cite{gerhardts91,transport}
have been performed in search of evidence
for a self-similar energy spectrum,
the so called Hofstadter butterfly.\cite{butter}
A main theme in the FIR absorption experiments has been to detect
and classify the rich variety of collective modes  that
occur in these systems.\cite{kern91,bollweg95,fir}

Along with the experimental work, theorists have addressed the
same
issues.\cite{wu93,huang93,gudmund96,gudmund98,petschel93,gudmund95,hall97}
The main difficulty lies in the fact that while the superlattice
is periodic, the Hamiltonian (including a vector potential) is not.
Most recent calculations of superlattice electronic structure
have used the Ferrari basis to deal with this matter.\cite{ferrari90}
We will instead apply ray-group-theoretical
techniques\cite{brown64,groups,hans82,hansetal}
to effectively reduce the calculational complexity.

With this approach, we are able to go beyond earlier calculations.
We find the band structure for {\em interacting} electrons in
general ``rational'' magnetic fields [i.e., the flux through 
a unit cell is $(L/N)\Phi_0$, where $\Phi_0$ is a flux quantum and 
$L, N \in \cal{Z} $]. 
Consequently, we are able to trace even fine-scale features of the
butterfly and at the same time study the effects screening has on it.
We also explore the FIR response.
The resulting spectra are rather rich. Along with absorption peaks caused by
collective modes, and known from experiments,\cite{kern91,bollweg95}
we find additional ones of mostly quantum-mechanical origin.

The antidot superlattice considered here is of simple square
symmetry ${\bf R}=n_1{\bf a}_1+n_2{\bf a}_2$, 
with lattice parameter $a$.
The  effective one-particle Hamiltonian is
\begin{equation}
\label{eq:ham}
  H = \frac{1}{2m} \left( {\bf p} + \frac{e}{c} {\bf A} \right)^2 
  + \sum_{\bf G} v({\bf G}) {\rm e}^{i{\bf G}\cdot{\bf r}},
\end{equation}
where the vector potential ${\bf A}={\bf B}\crossprod {\bf r}/2$ 
(symmetric gauge) describes the perpendicular magnetic field ${\bf B}$
and ${\bf G}=g_1{\bf b}_1+ g_2{\bf b}_2$, 
denotes the reciprocal lattice vectors.
We use GaAs parameters and work with short-period 
superlattices with
$a = 1000$ \AA{} and electron density $n_s = 1.2\cdot 10^{11}$ cm$^{-2}$.
A typical magnetic field $B = 1.65$ T gives four flux quanta 
per unit cell, filling factor $\nu=1.5$, and the cyclotron energy 
$\hbar \omega_c = 2.86$ meV, where $\omega_c = eB/mc$.
The last term in Eq.\ (\ref{eq:ham}) is a sum of the external
superlattice potential, described by a few principal Fourier
components,\cite{Note0} and the Hartree potential.
As for electron spin, we keep the twofold degeneracy in mind when
counting states, but neglect other effects such as Zeeman splitting and 
exchange interaction.

The electronic states we set out to solve for will satisfy 
the modified Bloch conditions
\begin{equation}
\label{eq:bloch}
  \hat{T}_M({\bf a}_{1(2)}) \psi_{\bf q} = 
  {\rm e}^{-i {\bf a}_{1(2)} \cdot {\bf q}} \psi_{\bf q}
\end{equation}
when the magnetic flux through a unit cell equals an
integer number of flux quanta $\Phi_0$.
Here
\begin{equation}
\label{eq:tmr}
  \hat{T}_M({\bf R}) = \exp \left[-\frac{i}{\hbar}{\bf R}\cdot
  \left( {\bf p}-\frac{e}{c}{\bf A} \right) \right]
\end{equation}
are magnetic translation operators forming a ray group,
\cite{hammer89} and 
the eigenstates can still be classified by different values of 
the crystal momentum
${\bf q} = q_1{\bf b}_1 + q_2{\bf b}_2$, in the first Brillouin zone.
\cite{brown64}
Actually, one 
finds sets of $L$ linearly independent functions each
transforming according to the same irreducible representation. 
This manifests itself as the splitting of the Landau band 
into $L$ subbands. 
For rational fields with
flux $(L/N)\Phi_0$ per unit cell, the irreducible representations of 
the group (\ref{eq:tmr}) are $N$-dimensional 
and the states $N$-fold degenerate.
This calls for a generalized treatment\cite{brown64,hans82,Note1}
which we have implemented but do not further describe here.

The next important step towards a solution is a canonical 
coordinate transformation.\cite{hans82}  We switch to
dimensionless units\cite{Note2} to be used hereafter, and introduce
\begin{equation}
\label{eq:strange}
  \xi (\eta) = \pm p_y + x/2, \quad p_{\xi (\eta)} =p_x \mp y/2.
\end{equation}
This preserves the canonical commutators, maps the 
kinetic energy of the Hamiltonian in the symmetric gauge
onto a harmonic oscillator in $\xi$, and makes the 
magnetotranslations act only on $\eta$,
\begin{equation}
\label{eq:separ}
  H_0 = \frac{1}{2} \left( \xi^2 + p_{\xi}^2\right),\,
  \hat{T}_M({\bf R}) = \exp \left( - i R_x p_{\eta} + i R_y \eta \right).
\end{equation}
The periodic potential mixes the $\xi$ and $\eta$ degrees of freedom
(in these coordinates it behaves like a magnetic translation operator)
\begin{eqnarray}
  H_1 &=& \sum_{\bf G} v({\bf G}) \hat{X}({\bf G}|\xi) 
  \hat{Y}({\bf G}|\eta), \nonumber\\
  &&\hat{X}({\bf G}|\xi)  = \exp(i G_x \xi  - i G_y p_{\xi}), \\
  &&\hat{Y}({\bf G}|\eta) = \exp(i G_x \eta + i G_y p_{\eta}). \nonumber
\end{eqnarray}
Using 
projection-operator techniques we find the symmetry-adapted
$\eta$-dependent functions
\begin{equation}
\label{eq:comb}
  \varphi({\bf q},l|\eta) = \sum_{m=-\infty}^{\infty}
  {\rm e}^{2\pi i m q_1} \delta \left(\eta + 
  \frac{a q_2}{L} - \frac{al}{L}-am \right),
\end{equation}
labeled by the magnetic crystal momentum ${\bf q}$ in the first 
magnetic Brillouin zone (i.e.,  $0 \leq q_1, q_2 < 1$) and 
the subband index $l = 0, \ldots, L-1$. 
Now the Ansatz 
\begin{eqnarray}
\label{eq:ansatz}
  \psi ({\bf q},l|\xi,\eta) &=& \sum_{l=0}^{L-1} 
  \varphi({\bf q},l|\eta) \sum_{n=0}^{\infty} a_{nl} \chi_n(\xi)
\end{eqnarray}
for the eigenstates
allows for subband mixing, 
and the $\xi$-dependence is accounted for by an expansion in harmonic 
oscillator eigenfunctions $\chi_n$.   Inserted into the Schr\"odinger
equation, the Ansatz yields the eigenvalue problem determining the electron
states
\begin{eqnarray}
\label{eq:eigen}
  \sum_{n' l'}&& \Big\{ \delta_{n n'}\delta_{l l'}
  \left( n + \frac{1}{2} \right) \nonumber \\
  &&+ \sum_{\bf G} v({\bf G}) 
  A_{l l'}({\bf G}) B_{n n'}({\bf G}) \Big\} a_{n' l'} = E a_{n l}.
\end{eqnarray}
The subband and Landau-level mixing coefficients are
\begin{eqnarray}
\label{eq:over}
  A_{l l'}({\bf G}) &=& {\rm e}^{2\pi i \left[ 
  g_1 g_2 /2 + g_1 l + (q_1 g_2 - q_2 g_1) + q_1 (l - l')
  \right] /L} \delta_{l',l+g_2}^{({\rm mod } L)},\nonumber \\
  B_{n n'}({\bf G}) &=& \int_{-\infty}^{\infty} \chi_n(\xi) 
  \hat{X}({\bf G}|\xi) \chi_{n'}(\xi) d\xi.
\end{eqnarray}
Equation (\ref{eq:eigen}) must be iterated together with
the Poisson equation updating the Hartree potential until 
self-consistency is reached.

\begin{figure}
  \centering
  \includegraphics[angle=270,width=0.5\textwidth]{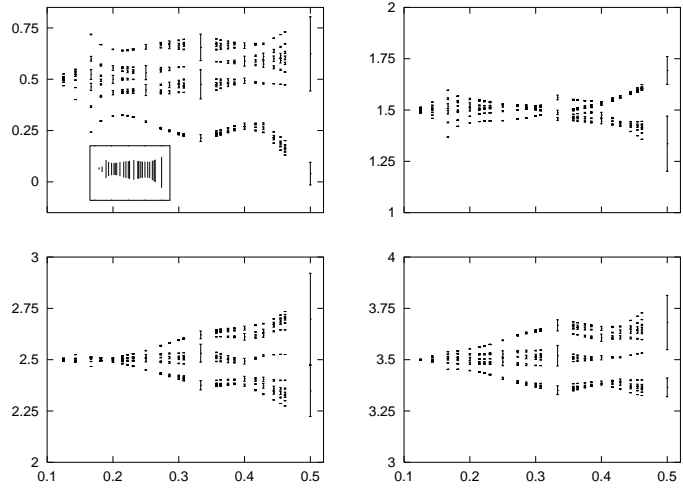}
  \vspace{5mm}
\caption{The width of the four lowest Landau bands 
(in units $\hbar \omega_c$) plotted versus inverse magnetic flux.
The four principal Fourier components of the potential 
[$v^{\rm ext}(0,\pm 1)$ and $v^{\rm ext}(\pm 1,0)$] are set 
to $1.43$ meV, $a = 1000$ {\AA}, and $n_s = 1.2\cdot10^{11}$ cm$^{-2}$.
The bands are centered around the
limiting Landau level values $n+1/2$ and get broader as the magnetic
field decreases. The commensurability phenomena manifest themselves
in the intricate splitting of the bands. To underscore the nonmonotonous 
dependence of broadening on the magnetic field, we also display the 
overall band widths in the inset of the left upper graph.
}
\label{fig_butter}
\end{figure}

Figure \ref{fig_butter} shows the splitting of the first 
four Landau levels as a function of the dimensionless inverse 
flux $\Phi_0/(Ba^2) = N / L$. 
With a reasonable computational effort we could treat
rational fields with $L \le 14$  and all possible 
$N$'s. This is enough to clearly resolve the intricate subband clustering.
\cite{butter}

It is easy to see that the bandwidths in Fig.\ \ref{fig_butter} 
decrease with increasing magnetic flux; however, the decrease  
is not monotonous. Instead they have maxima for flux values 6, 3, 
and 2, (see the inset) when there are 1, 2, and 3 completely 
filled Landau levels, respectively. Then the 2DES cannot screen 
the external potential very effectively, and the Fourier coefficients 
$v({\bf G})$ are larger than for other flux values.
Thus, since the bandwidth is set by a competition between the
band-narrowing effects of the magnetic field and the band broadening 
tendencies of the potential, this leads to a cusped behavior of the 
band top and bottom at integer filling factors. For the filling 
factors $\nu \le 1$ we also observe the same qualitative behavior 
while quantitative predictions of the Hartree theory
in this region may be inaccurate. We note that there exist other
(unrelated to electron-electron interaction) mechanisms
which also lead to nonmonotonous bandwidths.\cite{hall97}

Electron-electron interaction also contributes to diminishing
the symmetry of the butterfly as strong coupling
between different bands does.\cite{petschel93,hall97}
The second and third bands in Fig.\ \ref{fig_butter} which are
traversed by the chemical potential $\mu$ show reduced
regularity if compared to well-separated noninteracting  bands 
in Fig.\ 3 (a) of Ref.\ \onlinecite{hall97}, whereas our fourth band, 
well above $\mu$, would exhibit a considerable resemblance to 
Hofstadter's one-band picture when properly rescaled.

Turning to the dynamic response of the 2DES, we calculate
the density-density response function 
$ R_{{\bf G}{\bf G}'}({\bf k},\omega)$ 
within the random-phase approximation (RPA) by solving the set of equations
\begin{equation}
\label{eq:rpa}
  R_{{\bf G}{\bf G}'} = P_{{\bf G}{\bf G}'} +
  \sum_{{\bf G}''} P_{{\bf G}{\bf G}''}
  V_{\rm ee}({\bf k} + {\bf G}'') R_{{\bf G}''{\bf G}'}.
\end{equation}
Here $P_{{\bf G}{\bf G}'}$ is the independent particle 
response function which we can evaluate knowing the electron eigenstates
already calculated.
The FIR absorption 
of the long wavelength (${\bf k}$ in the first Brillouin zone)
light is proportional to 
$-\omega {\rm Im} R_{00}(\omega)$.\cite{dahl90}

\begin{figure}
  \centering
  \includegraphics[angle=0,width=0.9\textwidth]{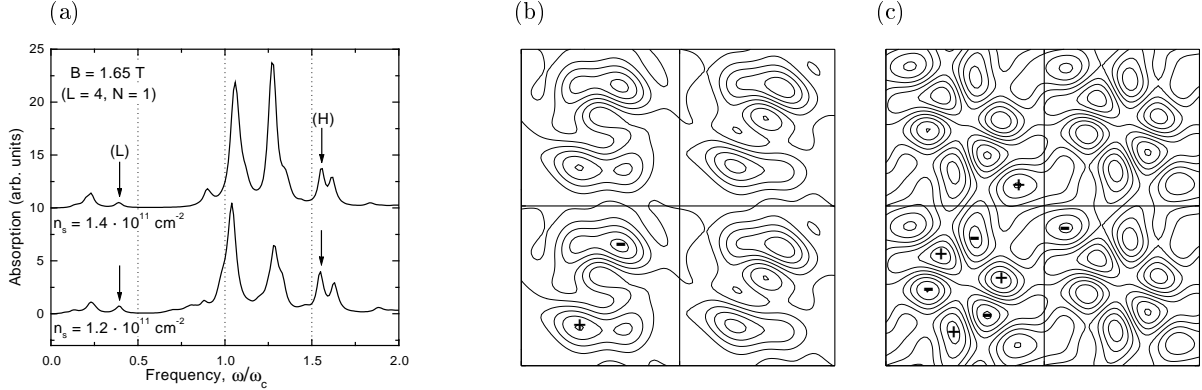}
  \vspace{5mm}
\caption{
Absorption spectra (panel a) and the oscillating charge-density distributions
associated with the peaks indicated by arrows 
(H $\leftrightarrow$ panel b) and (L $\leftrightarrow$ panel c),
respectively.    The upper curve in (a) is offset by $10$. 
The external potential has Fourier components
$v^{\rm ext}(0,\pm 1) = v^{\rm ext}(\pm 1,0) = \hbar\omega_c = 2.86$ meV
and
$v^{\rm ext}(0,\pm 2) = v^{\rm ext}(\pm 2,0) = \hbar\omega_c/4$.
For the high-frequency 
mode one sees in (b) a dipole rotating around the center of each lattice cell.
For the low-frequency 
mode one can in (c) observe a dipole centered on each of the antidots and a 
hexagonal pattern between the antidots. One structure of each kind is marked 
with signs. 
}
\label{fig_fir}
\end{figure}

The so calculated spectra typically exhibit several
conspicuous peaks. 
In Fig.\ \ref{fig_fir} (a), we display spectra calculated for
electron densities $n_s = 1.2\cdot10^{11}$ cm$^{-2}$ and 
$1.4\cdot10^{11}$ cm$^{-2}$, respectively, 
and wave vector ${\bf k}= (\pi/10)a^{-1}\hat{x}$.
Following the suggestion of Ref.\ \onlinecite{gudmund98} to classify
the different peaks by studying the corresponding charge fluctuations;
we also trace their development in time to pick out the ones that are
stable with respect to changing electron density. Here we try to 
concentrate on a few of these plots and give a thorough discussion of 
them. To this end we calculate the time-dependent, induced charge 
density at the absorption-peak frequencies in four adjacent unit cells, 
and plot snapshots thereof in Fig.\ \ref{fig_fir} (b) and (c). 
The antidots are situated at the intersections of the thick lines 
and the ``+''(``-'') signs mark the locations of the charge density 
maxima (minima). 

The two spectra shown in Fig.\ \ref{fig_fir} (a)
calculated for different electron densities are very similar.
At the same time, however, the induced charge densities 
at the different peaks can
in general change quite a lot with changing electron density.
There are a few exceptions to this, most notably the peaks marked
(H) and (L) and indicated by arrows. 
As we will see, one can give a clear, semiclassical interpretation to
these modes.

Thus, Fig.\ \ref{fig_fir} (b) shows the charge density
corresponding to the (H) peak in Fig.\ \ref{fig_fir} (a). 
One sees a dipole which, looking at a sequence of snapshots,
rotates around the center of each lattice
cell (i.e., between four antidots) in the direction of 
cyclotron motion. 
This mode, which can be  anticipated on general grounds, 
emerges in simple theoretical models\cite{wu93} and
has been detected experimentally.\cite{kern91,bollweg95}

In the low frequency region we find a more complicated
collective mode [peak (L) in Fig.\ \ref{fig_fir} (a)] depicted
in Fig.\ \ref{fig_fir} (c). 
A dipolar charge distribution is
rotating around each antidot {\em in} the direction of 
cyclotron motion, and a ``ring'' of three charge density 
maxima and three minima between the antidots is rotating in the
{\em opposite} direction. During each period one sees some small 
charge transfer between the two structures.
We interpret this mode as a pair of coupled
(inter)edge magnetoplasmons
with angular momenta $l=+1$ and $l=-3$ around an antidot
and the center of a cell, respectively. 
The dynamics of this mode is mainly determined by an equilibrium
between the Lorentz force and restoring electrostatic forces.  From 
this follows that the magnetoplasmon propagates in opposite
directions around a charge density maximum (a cell center) and a minimum 
(an antidot).\cite{sommer95}
Note also that this mode is an example of mixing of states 
with angular momenta differing by a multiple of four 
in a square lattice.

Both peaks, (H) and (L), show absorption of light
polarized in the direction of the cyclotron resonance in
agreement with experiment.\cite{bollweg95} The
existing theoretical explanation,\cite{wu93} however, 
is based on a model with circularly symmetric unit cells and cannot
describe the interplay of modes centered at different
places of the unit cell. 
Besides the modes discussed until now the 2DES absorbs energy at a number
of other frequencies. The corresponding induced charge distributions are 
more complex than the ones displayed in Figs.\ \ref{fig_fir} 
(b) and (c).
These excitations are to a large extent of a quantum-mechanical nature,
i.e., the result of intersubband and inter-Landau-level transitions
(see also Ref.\ \onlinecite{gudmund96}).
In this context, it is also clear that our spectra obtained for
short-period superlattices are not completely comparable to
the ones found experimentally.    There are two main reasons for this,
the potential has a stronger influence on electron motion, and we have not 
treated disorder broadening.

In conclusion, we have developed a theory that makes it possible to study
the electronic structure 
of a 2DES in a combined periodic potential and strong magnetic field
in a detailed fashion 
treating electron-electron interactions at the mean-field level.
The so calculated energy spectra show clear traces of a self-similar
structure like the Hofstadter butterfly, however,
considerably modified by screening effects.
The dielectric response of the 2DES in the FIR regime reveals a rich 
spectrum of excitations.   
Some of the peaks in these spectra can be interpreted in terms of
semiclassical collective excitations, 
while others mainly are of a quantum-mechanical origin.

We thank Carlo Canali, Carl-Olof Almbladh, and Koung-An Chao 
for valuable discussions.
One of us (P.J.) is supported by 
the Swedish Natural Science Research Council (NFR). 


\end{multicols}

\hrule


\vspace{10mm}

\vspace{10mm}

\end{document}